\documentstyle[aps,multicol,epsf]{revtex}

\begin{document}
\draft
\title{Evolution of reference networks with aging}

\author{S.N. Dorogovtsev$^{1, 2, \ast}$ and J.F.F. Mendes$^{1,\dagger}$}

\address{
$^{1}$ Departamento de F\'\i sica and Centro de F\'\i sica do Porto, Faculdade de Ci\^encias, 
Universidade do Porto\\
Rua do Campo Alegre 687, 4169-007 Porto -- Portugal\\
$^{2}$ A.F. Ioffe Physico-Technical Institute, 194021 St. Petersburg, Russia 
}

\maketitle

\begin{abstract}
We study the growth of a reference network with aging of sites defined in the following way. Each new site of the network is connected to some old site with probability proportional (i) to the connectivity of the old site as in the Barab\'{a}si-Albert's model and (ii) to $\tau^{-\alpha}$, where $\tau$ is the age of the old site. We consider $\alpha$ of any sign although reasonable values are 
$0 \leq \alpha \leq \infty$.
We find both from simulation and analytically that the network shows scaling behavior only in the region $\alpha < 1$. When $\alpha$ increases from $-\infty$ to $0$, the exponent $\gamma$ of the distribution of connectivities ($P(k) \propto k^{-\gamma}$ for large $k$) grows from $2$ to the value for the network without aging, i.e. to $3$ for the Barab\'{a}si-Albert's model. The following increase of $\alpha$ to $1$ makes $\gamma$ to grow to $\infty$.
For $\alpha>1$ the distribution $P(k)$ is exponentional, and the network has a chain structure. 
\end{abstract}

\pacs{PACS numbers: 05.10.-a, 05.40.-a, 05.50.+q, 87.18.Sn}

\begin{multicols}{2}

\narrowtext


 The explosion of the general interest on the problem of the structure and evolution of most different networks \cite{lah,red,wat,bar1,hub,bar2,watbook} 
is connected not only with the sudden understanding that our world is in fact a huge set of various networks (with the most striking examples of Web and neural networks 
 -- see the papers \cite{bar1,hub,par,mon} and references therein) but also with the recent finding that 
many networks obey scaling laws \cite{red,bar2}. 
That was the reason to renew old studies \cite{erd,bol,der,sta}. 
One may note also that the growth of networks is only a particular kind of fascinating growth processes \cite{barbook,kar,kim}.   

The simplest kind of growing networks is the reference networks \cite{ref} in which new links appear only between new sites and the old ones but they never appear between old sites. Therefore, in the reference networks, the average connectivity of old sites is always higher than that one of younger sites. The well-known example of the reference networks is the network 
of citations of scientific papers \cite{red} in which each paper is a site of the corresponding net and links are the references to the cited papers. 

A clear beautiful model of a reference network that shows 
scaling behavior was recently proposed by Barab\'{a}si and Albert \cite{bar2}. In their network, each new site is connected with some old site with 
probability proportional to its connectivity $k$. In this case, the distribution of the connectivities in the large network (i.e. one of the most considerable characteristics of the structure and evolution of networks) 
shows 
a power-law dependence $P(k) \propto k^{-\gamma}$ with the exponent $\gamma=3$. 

However, in real reference networks aging usually occurs: alas, we rarely cite old papers!   
One may ask, how do the structure of the network change if the aging of sites is introduced, i.e. if the probability of connection of the new site with some old one is proportional not only to the connectivity of the old site but also to the power of its age, $\tau^{-\alpha}$, for example? This question is quite reasonable: indeed, Fig. 1 demonstrates that the structure of the network depends obviously on $\alpha$.
We show below both numerically and analytically that this change is dramatic: the scaling disappears when $\alpha>1$, and the exponents of the scaling laws depend strongly on $\alpha$ in the range $\alpha<1$. This result could not be foreseen beforehand: the scaling is very sensitive to changes of the model. For example, as it is noted in \cite{bar2}, the scaling disappears if the probability of the connection to an old site is proportional to $k^\epsilon, \epsilon \neq 1$. 

Here, we consider exclusively the network, in which only one extra link appears when the new site is added,
since the results for the exponents do not depend on the number of links which are added each time 
\cite{bar2}. 

Let us start from the simulation which turns to be easy for the problem under consideration because we study only the characteristics dependent on the connectivity. They are: (i) the distribution of connectivities in the network $P(k,t)$ in the instant $t$ (only one site is in the network at $t=0$, and one more site is added in each instant) and (ii) the mean connectivity $\overline{k}(s,t)$ of the site $s$ ($0<s<t$) in the instant $t$, i. e. the local density of the connectivities. If one is interested only in these quantities, there is no need to keep a matrix of connections in memory, and the simulation is very fast. 

The results of the numerics are shown in Figs. 2-5. Although only 
the region $\alpha \geq 0$ seems to be of real significance, we consider also negative values of $\alpha$ since they do not lead to any contradiction. 
One may see clearly from the figures that $P(k,t) \propto k^{-\gamma}$ for large $k$ and $\overline{k}(s,t) \propto s^{-\beta}$ for small $s$, where $\beta$ is the other scaling exponent, only for $\alpha<1$. As it should be, we get the values $\gamma=3$ and $\beta=1/2$ for $\alpha=0$ \cite{bar2}. 
Note that, at $0<\alpha<1$, the deviations of the dependence $\log P(k)$ vs. $\log k$ for small $k$
are stronger as in the real reference network \cite{red}, than those ones at $\alpha=0$.
At $\alpha>1$, $P(k)$ turns to be exponential, and the mean connectivity 
tends to be constant at large $s$. Thus, while $\alpha$ changes from $0$ to $1$, 
$\gamma$ grows from $3$ to $\infty$, and $\beta$ decreases from $1/2$ to $0$. One sees from Fig. 3, that, for positive $\alpha$ the $\log P(k)$ vs. $\log k$ dependence looks more curved than those ones for $\alpha=0$ like in the real reference network \cite{red}.

The simulations demonstrate also that $\gamma$ has a tendency to decrease from $3$ to $2$, and $\beta$ 
to grow from $1/2$ to $1$ when $\alpha$ decreases from $0$ to $-\infty$. 

Let us show now how these results may be described. We use an effective medium approach, inspired by   \cite{bar2}. We shall explain why it gives so good results for the exponents later but first we outline the applied theory for the problem under consideration. 

The main {\it ansatz} of the effective medium approach in our case is the approximation of the probability $P(k,s,t)$ that the connectivity of the site $s$ at the moment $t$ is equal to $k$ by the $\delta$-function:    

\begin{equation}
\label{1}
P(k,s,t) = \delta(k - \overline{k}(s,t))
.
\end{equation}
Thus, the exact function $P(k,s,t)$ is assumed to be enough sharp. We suppose also that we may use a continuous approximation. These two strong assumptions lead immediately to the following equation:

\begin{equation}
\label{2}
\frac{\partial \overline{k}(s,t)}{\partial t} = 
\frac{\overline{k}(s,t)(t-s)^{-\alpha}}{\int\limits_0^t du \overline{k}(u,t)(t-u)^{-\alpha}} \
,  \ \overline{k}(t,t)  =  1 
.
\end{equation}
Here, the boundary condition means that only one link is added each time. 

One may check that Eq. (\ref{2}) is consistent. Let us apply $\int_ 0^t ds$ to it. Then,

\begin{equation}
\label{3}
\int_ 0^t ds\frac{\partial \overline{k}(s,t)}{\partial t} =
\frac{\partial }{\partial t} \int_ 0^t ds \overline{k}(s,t) - \overline{k}(t,t) = 1
,
\end{equation}
and we get immediately the proper relation

\begin{equation}
\label{4}
\int_ 0^t ds \overline{k}(s,t) = 2t 
,
\end{equation}
i. e. the sum of connectivities equals the doubled number of links in the network.

We search for the solution of Eq. (\ref{2}) in the scaling form
\begin{equation}
\label{5}
\overline{k}(s,t) \equiv \kappa(s/t) \ , \ s/t \equiv \xi 
,
\end{equation}
which is consistent also with Eq. (\ref{4}). Then  Eq. (\ref{2}) becomes 

\begin{eqnarray}
\label{6}
-\xi (1-\xi)^\alpha \frac{d \ln \kappa(\xi)}{d \xi}  & = & 
\left[ \int_0^1 d\zeta \kappa(\zeta) (1-\zeta)^{-\alpha} \right]^{-1} \equiv \beta 
, 
\nonumber \\[3pt]
\kappa(1) = 1
,
\end{eqnarray}
where $\beta$ is a constant, which is unknown yet. We shall see soon that this constant is the exponent of the mean connectivity. Eqs. (\ref{4}) and (\ref{5}) give the relation
$\int_0^1 d\zeta \kappa(\zeta) = 2$.

The solution of Eq. (\ref{6}) is 
\begin{equation}
\label{7}
\kappa(\xi) = B \exp \left[-\beta \int \frac{d\xi}{\xi(1-\xi)^\alpha}  \right]
,
\end{equation}
where $B$ is a constant. The indefinite integral in Eq. (\ref{7}) may be taken:

\begin{eqnarray}
\label{8}
\int \frac{d\xi}{\xi(1-\xi)^\alpha} = \phantom{WWwwWWWWWWWWWW}
\nonumber \\
\ln \xi + \sum_{k=0}^{\infty} \frac{1}{k!(k+1)^2} 
\alpha(\alpha+1) \ldots (\alpha+k)\xi^{k+1}  =
\nonumber \\ [2pt]
\ln \xi + \alpha\, _3 F_2(1,1,1+\alpha;2,2;\xi)
,\phantom{WWWWWWw}
\end{eqnarray}
where $_3 F_2(\ ,\ ,\ ;\ ,\ ;\ )$ is the hypergeometric function \cite{prubook}.
Recalling the boundary condition $\kappa(1) = 1$, we find the constant $B$. Thus the solution is

\begin{eqnarray}
\label{9}
\kappa(\xi) = \phantom{WWwwWWWeWWWWWWWWWWWW}
\nonumber \\[3pt]
e^{-\beta(C + \psi(1-\alpha))} \xi^{-\beta} \exp 
\left[ -\beta \alpha \xi\, _3 F_2(1,1,1+\alpha;2,2;\xi)  \right]
,
\end{eqnarray}
where $C=0.5772 \ldots$ is the Euler's constant and $\psi(\ )$ is the $\psi$-function. 
Now we see that the constant $\beta$ indeed is the exponent of mean connectivity, 
since $\kappa(\xi) \sim \xi^{-\beta}$ if $\xi \to 0$. The transcendental equation for $\beta$ may be written 
if one substitutes 
Eq. (\ref{9}) into the right side of Eq. (\ref{6}):

\begin{eqnarray}
\label{10}
\beta^{-1} =  e^{-\beta(C + \psi(1-\alpha))} \times     \phantom{WWWWWWWWWWW}
\nonumber \\[3pt]
\int_0^1 \frac{d\zeta}{\zeta^{\beta} (1-\zeta)^{\alpha}} \exp 
\left[ -\beta \alpha \zeta\, _3 F_2(1,1,1+\alpha;2,2;\zeta)  \right]
,
\end{eqnarray}
that is our main equation. 

Before we shall find the solution of Eq. (\ref{10}), let us show how the exponents $\beta$ and $\gamma$ are related if one applies Eq. (\ref{1}) of the effective medium approach.
$\overline{k}(s,t) \propto s^{-\beta}$, so $s \propto k^{-1/\beta}$. Hence,
$k^{-\gamma} \propto P(k,t) \propto \partial s/\partial k \propto k^{-1-1/\beta}$, and one obtains that

\begin{equation}
\label{11}
\gamma = 1 + 1/\beta
.
\end{equation}

The solution of Eq. (\ref{10}) exists in the range $-\infty < \alpha < 1$. The results of the numerical solution are shown in Fig. 5. One may also find $\beta(\alpha)$ and $\gamma(\alpha)$ 
at $\alpha \to 0$: 

\begin{equation}
\label{12}
\beta \cong \frac{1}{2} - (1 - \ln 2)\alpha \ , \ 
\gamma \cong 3 + 4(1 - \ln 2)\alpha
,
\end{equation}
where the numerical values of the coefficients are 
$1 - \ln 2 = 0.3069\ldots$ and $4(1 - \ln 2) = 1.2274\ldots$.
We used the relation
$_3F_2(1,1,1;2,2;\zeta) = Li_2(\zeta)/\zeta \equiv 
(\sum_{k=1}^\infty \zeta^k/k^2)/\zeta$
while deriving Eq. (\ref{12}). Here, $Li_2(\ )$ is the polylogarithm function of order 2 
\cite{prubook}.

In the limit of $\alpha \to 1$, using the relation
$_3F_2(1,1,2;2,2;\zeta) = -\ln(1-\zeta)/\zeta$, we find

\begin{equation}
\label{13}
\beta \cong c_1 (1-\alpha) \ , \ 
\gamma \cong \frac{1}{c_1} \frac{1}{1-\alpha}
.
\end{equation}
Here,
$c_1=0.8065\ldots,c_1^{-1}=1.2400\ldots$: the constant $c_1$ is the solution of the equation
$1+1/c_1 = \exp(c_1)$. Note also that $\beta \to 1$ and $\gamma \to 2$ 
in the limit $\alpha \to -\infty$.

One sees from Fig. 5 that the results of the simulation and of the analytical calculations are in qualitative correspondence. Deviations may be noticed only in regions where simulations can not provide sufficient precision. Why does such a coarse theory demonstrate so close agreement with the simulation? One may show \cite{dms} that the function $P(k,s,t)$ of $k$ is functionally very sharp as compared with the studied long-tailed distribution $P(k,s)$. Therefore, the main {\it anzats} Eq. (\ref{1}) gives excellent results. 
In fact, Eqs. (\ref{1}) and (\ref{2}) provides us easily with all known results on reference networks.

One may imagine from Fig. 1 that the point $\alpha=1$ ($\beta \to 0$ and $\gamma \to \infty$) at which the scaling collapses marks the transition from the multidimensional network for $\alpha<1$ to the chain structure. Fig. 1 demonstrates also that, in the case of $\alpha \to -\infty$ ($\beta \to 1$ and $\gamma \to 2$) all sites of the network are connected with the oldest one. 
The behavior of the considered quantities near these points evokes associations with the lower and higher critical dimensions in the theory of usual phase transitions \cite{amibook}.

There are some possibilities to change the exponents of the network without aging 
\cite{bar2,dms}. One may check that, in these cases, the range of variation of the exponents 
 $\beta$ and $\gamma$ when $\alpha$ changes from $-\infty$ to $1$ is the same as in the present Letter. 

In summary, we have considered the reference network with the power-law ($\tau^{-\alpha}$) aging of sites. We have found both from our simulations and using the effective medium approach that the network shows scaling behavior only in the region $\alpha < 1$. 
We have calculated the exponents of the power-law dependences of the distribution of connectivities $P(k,t) \propto k^{-\gamma}$ and of mean connectivity $\overline{k}(s,t) \propto s^{-\beta}$ and have shown that they depend crucially on $\alpha$.

The following questions remain open. Is our unproved idea about the nature of the threshold point $\alpha=1$ reasonable? Do the analogies with the lower and higher critical dimensions exist indeed? 
What other quantities of the network demonstrates the scaling behavior?
\\

SND thanks PRAXIS XXI (Portugal) for a research grant PRAXIS XXI/BCC/16418/98. JFFM was partially supported by the projects PRAXIS/2/2.1/FIS/299/94. We also thank E. J. S. Lage for reading the manuscript and A. V. Goltsev, S. Redner, and A. N. Samukhin for many useful discussions.\\
$^{\ast}$      Electronic address: sdorogov@fc.up.pt\\
$^{\dagger}$   Electronic address: jfmendes@fc.up.pt

\begin{figure}
\epsfxsize=70mm
\epsffile{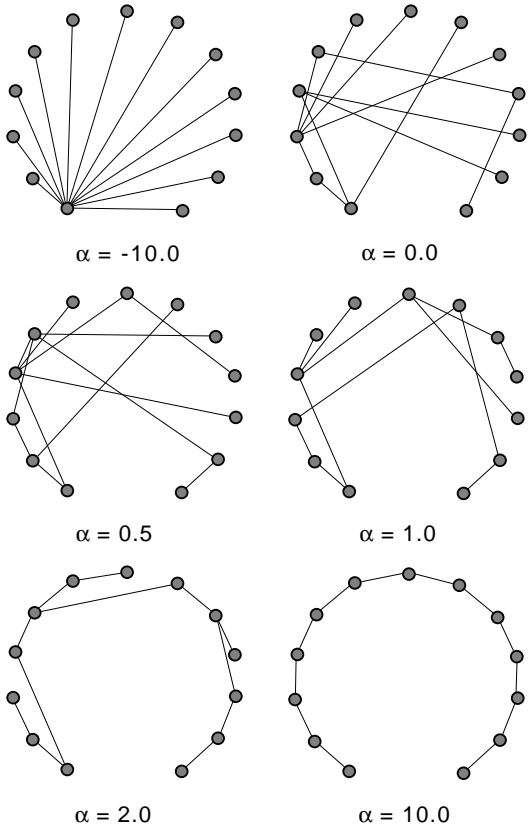}
\caption{
Change of the network structure with increase of the aging exponent $\alpha$. 
The aging is proportional to $\tau^{-\alpha}$, where $\tau$ is the age of the site. 
The network grows clockwise starting from the site below on the left. Each time 
one new site with one link is added.
}
\label{f1}
\end{figure}

\begin{figure}
\epsfxsize=75mm
\epsffile{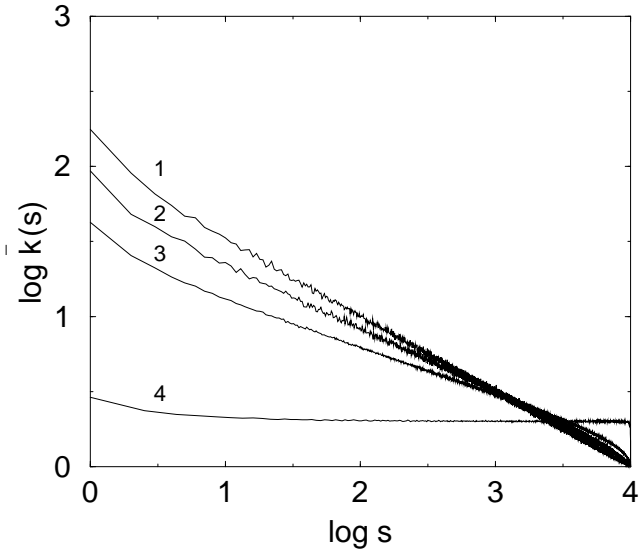}
\caption{
Mean connectivity vs. number $s$ of the site for several values of the aging exponent $\alpha$: 
1) $\alpha = 0$, 2) $\alpha = 0.25$, 3) $\alpha = 0.5$, 4) $\alpha = 2.0$.  
The network size is $t=10\,000$.
}
\label{f2}
\end{figure}

\begin{figure}
\epsfxsize=85mm
\epsffile{ 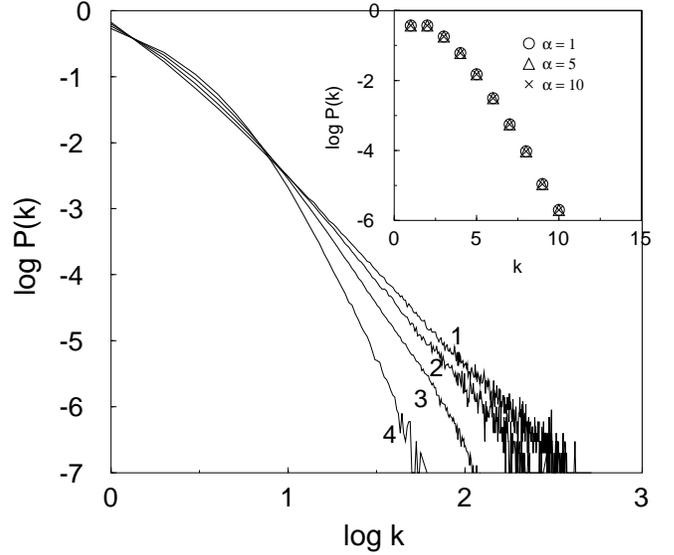}
\caption{
Distribution of the connectivities for several values of the aging exponent: 1) $\alpha = 0$, 2) $\alpha = 0.25$, 3) $\alpha = 0.5$, 4) $\alpha = 0.75$. The inset shows $\log P(k)$ vs. $k$ for $\alpha=1,5$, and $10$. Note that, in the later case, all three curves are nearly the same.
}
\label{f3}
\end{figure}

\begin{figure}
\epsfxsize=75mm
\epsffile{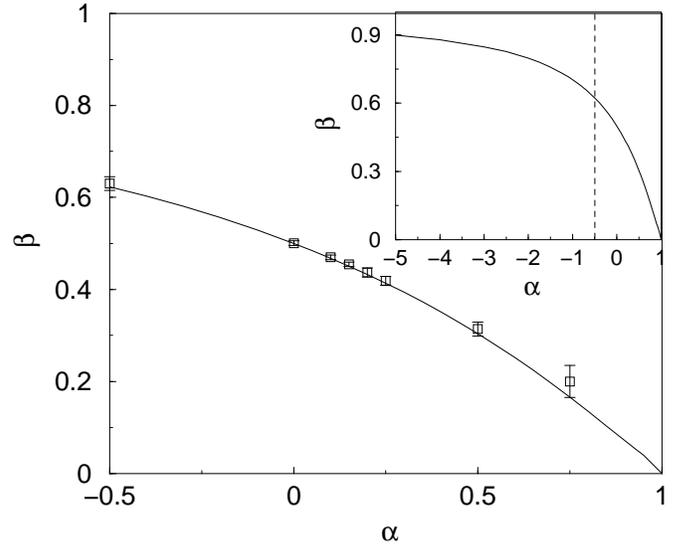}
\caption{
Exponent  $\beta$ of the mean connectivity vs. aging exponent $\alpha$. Points are obtained from the simulations. The line is the solution 
of Eq. (\protect\ref{10}). The inset shows the analytical solution in the range $-5<\alpha<1$. 
Note that $\beta \to 1$ if $\alpha \to -\infty$.
}
\label{f4}
\end{figure}

\begin{figure}
\epsfxsize=75mm
\epsffile{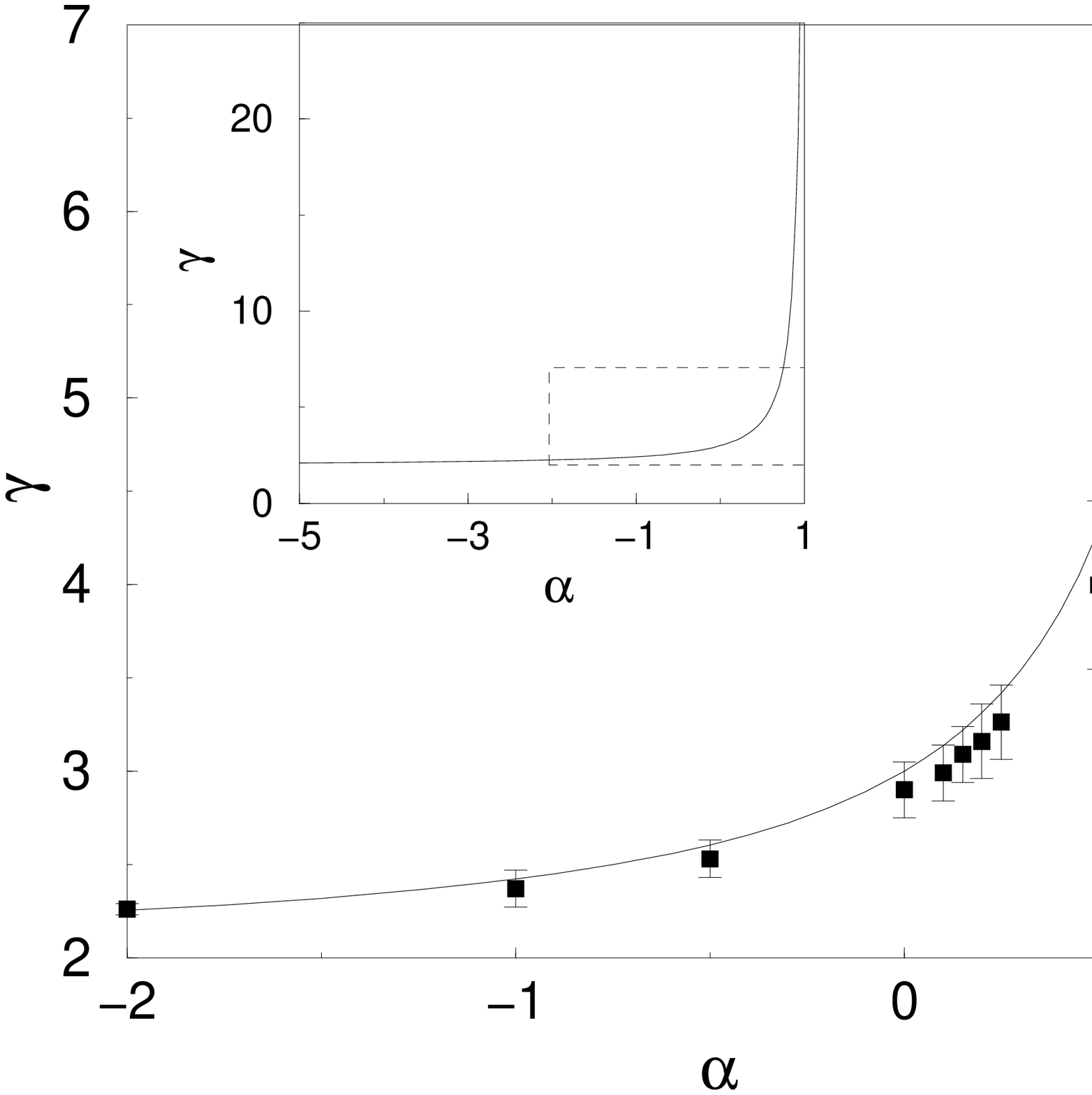}
\caption{
Exponent  $\gamma$ of the connectivity distribution vs. aging exponent $\alpha$. The points show the results of the simulations. The line is the solution of Eq. (\protect\ref{10}) with account for Eq. (\protect\ref{11}). The inset depicts the analytical solution in the range $-5<\alpha<1$.
}
\label{f5}
\end{figure}

\end{multicols}


\begin{references}

\bibitem{lah} J. Lahererre and D. Sornette, Eur. Phys. J. B {\bf 2}, 525 (1998).
\bibitem{red} S. Redner, Eur. Phys. J. B {\bf 4}, 131 (1998).
\bibitem{wat} D. J. Watts and S. H. Strogatz, Nature {\bf 393}, 440 (1998).
\bibitem{bar1} R. Albert, H. Jeong, and A.-L. Barab\'{a}si, Nature {\bf 401}, 130 (1999).
\bibitem{hub} B. A. Huberman, P. L. T. Pirolli, J. E. Pitkow, and  R. J. Lukose, Science {\bf 280}, 95 (1998).
\bibitem{bar2} A.-L. Barab\'{a}si and R. Albert, Science {\bf 286}, 509 (1999);  A.-L. Barab\'{a}si, 
R. Albert and  H. Jeong, Physica A {\bf 272}, 173 (1999).
\bibitem{watbook} D. J. Watts, {\it Small Worlds} (Princeton University Press, Princeton, New Jersey, 1999).
\bibitem{par} G. Parisi, J. Phys. A {\bf 19}, L675 (1986).
\bibitem{mon} R. Monasson and R. Zecchina, Phys. Rev. Lett. {\bf 75}, 2432 (1995).
\bibitem{erd} P. Erd\"{o}s and A. Ren\'{y}i, Publ. Math. Inst. Hung. Acad. Sci. {\bf 5}, 17 (1960).
\bibitem{bol} B. Bollob\'{a}s, {\it Random Graphs} (Academic Press, London, 1985).
\bibitem{der} B. Derrida and D. Stauffer, Europhys. Lett. {\bf 2}, 739 (1987).
\bibitem{sta} D. Stauffer, Philos. Mag. B {\bf 56}, 901 (1987).
\bibitem{barbook} A.-L. Barab\'{a}si, H. E. Stanley, {\it Fractal Concepts of Surface Growth} (Cambridge University Press, Cambridge, 1995).
\bibitem{kar} M. Kardar, G. Parisi, and Y. C. Zhang, Phys. Rev. Lett. {\bf 56}, 889 (1986).
\bibitem{kim} J. M. Kim and J. M. Kosterlitz, Phys. Rev. Lett. {\bf 62}, 2289 (1989).
\bibitem{ref} We are not sure that the term ``reference network'' is already accepted but it seems to be reasonable.
\bibitem{prubook} A. P. Prudnikov, Yu. A. Brychkov, and O. I. Marichev, {\it Integrals and Series}, vol. 1 (Gordon and Breach Science Publishers, New York, 1986).
\bibitem{dms} S.N. Dorogovtsev, J.F.F. Mendes, and A. N. Samukhin, to be published.
\bibitem{amibook} D. J. Amit, {\it Field Theory, the Renormalization Group, and Critical Phenomena} (McGraw-Hill, New York, 1978).



\end{references}
\end{document}